\PassOptionsToPackage{table,dvipsnames}{xcolor}

\documentclass[%
 aip,
 amsmath,amssymb,
 reprint,%
]{revtex4-1}

\usepackage{graphicx}
\usepackage{dcolumn}
\usepackage{bm}

\usepackage[utf8]{inputenc}
\usepackage[T1]{fontenc}
\usepackage{mathptmx}
\usepackage{etoolbox}

\usepackage{graphicx}
\usepackage{dcolumn}
\usepackage{bm}
\usepackage{xfrac} 
\usepackage{ulem} 
\usepackage{cancel}
\usepackage{mathtools}
\usepackage{url}
\usepackage{circuitikz}
\usepackage{tikz}
\usepackage{amsmath}
\usepackage{amssymb} 
\usepackage{multirow}
\usepackage{booktabs}
\usetikzlibrary{shapes,arrows,positioning,calc}
\usepackage{subcaption}

\usepackage{booktabs} 
\usepackage{tabularx} 

\usepackage{natbib}

\newcommand{\tauHK}{$\tau$HK }
\newcommand{\itwoc}{I$^2$C }
\newcommand\fallingedge{%
\begin{tikzpicture}[scale=0.3\baselineskip/18pt]
 \draw (0,1) -- (1,1) -- (1,0) -- (2,0);
\end{tikzpicture}%
}

\makeatletter
\def\@email#1#2{%
 \endgroup
 \patchcmd{\titleblock@produce}
  {\frontmatter@RRAPformat}
  {\frontmatter@RRAPformat{\produce@RRAP{*#1\href{mailto:#2}{#2}}}\frontmatter@RRAPformat}
  {}{}
}%
\makeatother
\begin{document}

\preprint{AIP/123-QED}


\title[$\tau$HK]{$\tau$HK: a modular housekeeping system for cryostats and balloon payloads}

\author{Simon Tartakovsky}
\affiliation{Department of Physics, Princeton University, Jadwin Hall, Princeton, NJ 08544, USA}
\author{Steven J. Benton}
\affiliation{Department of Physics, Princeton University, Jadwin Hall, Princeton, NJ 08544, USA}
\author{Aurelien A. Fraisse}
\affiliation{Department of Physics, Princeton University, Jadwin Hall, Princeton, NJ 08544, USA}
\author{William C. Jones}
\affiliation{Department of Physics, Princeton University, Jadwin Hall, Princeton, NJ 08544, USA}
\author{Jared L. May}
\affiliation{Department of Physics, Case Western Reserve University, 10900 Euclid Ave, Cleveland, OH 44106, USA}
\author{Johanna M. Nagy}
\affiliation{Department of Physics, Case Western Reserve University, 10900 Euclid Ave, Cleveland, OH 44106, USA}
\author{Ricardo R. Rodriguez}
\affiliation{Department of Physics, Case Western Reserve University, 10900 Euclid Ave, Cleveland, OH 44106, USA}
\author{Philippe Voyer}
\affiliation{Department of Physics, Princeton University, Jadwin Hall, Princeton, NJ 08544, USA}
\affiliation{Department of Mechanical \& Aerospace Engineering, Princeton University, Engineering Quad, Princeton, NJ 08544, USA}

\date{\today}

\begin{abstract}
\tauHK is a versatile experiment housekeeping (HK) system designed to perform cryogenic temperature readout and heater control on the upcoming Taurus balloon experiment\cite{may2024instrumentoverviewtaurusballoonborne}. $\tau$HK, more broadly, is also suitable for ambient-temperature applications and general-purpose experiment input and output. It is built around an IEEE Eurocard subrack capable of housing up to 16 interchangeable daughter cards, allowing a fully populated system to support as many as 256 independent channels while drawing under 7.5\,W. This modular architecture allows experiments to expand on the existing daughter cards with ones tailored to their specific needs. There are currently three flavors of daughter cards: Resistive Temperature Device (RTD) readout, general purpose thermometer bias and readout, and load driver. The RTD board consists of a low noise lock-in amplifier that is limited only by device sensitivity over all temperature ranges. The general-purpose bias and readout board with chopping capability is primarily designed for thermometer diodes, but flexible enough to accommodate room temperature thermistors, Wheatstone bridges, optical encoders, and other devices. Finally, the load driver card can output an analog voltage for precise cryogenic heaters or it can be used to pulse width modulate high power loads. \tauHK is a power efficient solution for experimental housekeeping needs that is suited for the the harsh environment of stratospheric ballooning.

\end{abstract}

\maketitle

\section{Introduction}

Many experimental fields require cryogenic cooling to achieve the desired performance (e.g.  \citet{TIMoverview} and \citet{may2024instrumentoverviewtaurusballoonborne}). As experiments grow in complexity, so do the cryostats supporting them, often requiring hundreds of thermometers and heaters to monitor and control performance \cite{tartakovsky2024thermalarchitecturecryogenicsuperpressure, spider_thermal_performance_jon}.

\textit{Lake Shore Cryotronics} provides one of the few commercially available solutions capable of measuring cryogenic temperature sensors such as resistive temperature devices (RTDs) and diodes. 
However, these commercial solutions do not scale efficiently for large cryogenic systems and are not suitable for the harsh environments of scientific ballooning. 
A bespoke housekeeping solution such as the BLASTbus presented in \citet{blastbus} is typically used, but by virtue of being specialized, such solutions are not easily adapted to future projects.
Large experiments, cryogenic and otherwise, would benefit from a unified system that can provide flexible analog and digital inputs and outputs (I/O). 

\begin{figure}
    \centering
    \includegraphics[width=\linewidth]{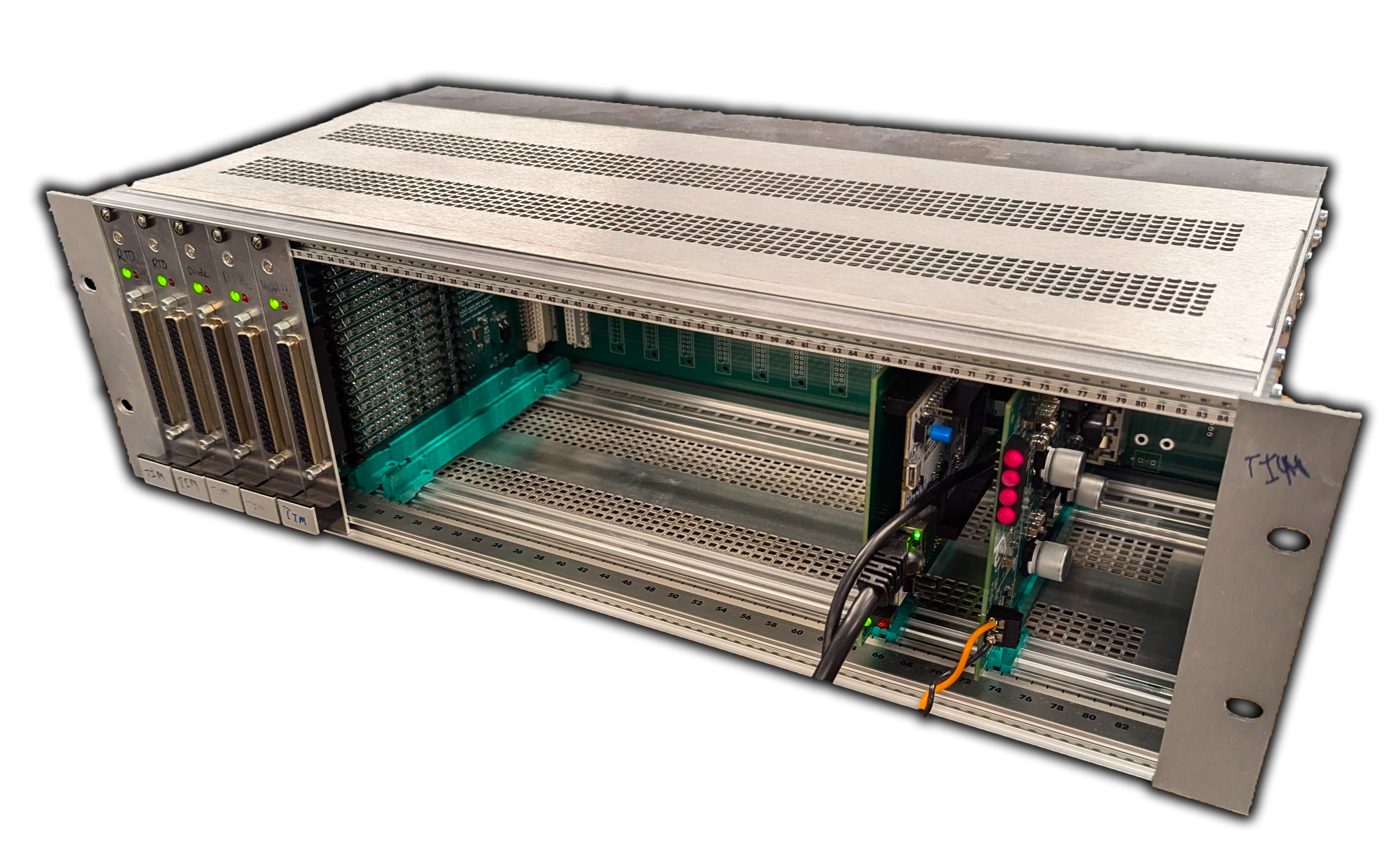}
    \caption{The \tauHK system used by the Terahertz Intensity Mapper (TIM) balloon experiment \cite{TIMoverview}. This system is populated with two RTD cards, two diode cards and one heater card.}
    \label{fig:picture}
\end{figure}

$\tau$HK (shown in Fig. \ref{fig:picture}) is a modular housekeeping system designed to support large cryogenic experiments, but remains flexible and cost-effective for general-purpose I/O on any large experimental platform.
Its modularity is achieved through interchangeable daughter cards tailored for specific readout use-cases.
A system overview is presented in Table \ref{tab:overview}.

\begin{table*}
    \centering
    \renewcommand{\arraystretch}{1.3} 
    \begin{tabularx}{0.8\textwidth}{|l|c|X|}
        \hline
        \rowcolor{gray!20} \textbf{$\tau$HK System} & \textbf{Value} & \textbf{Description} \\
        \hline
        Form factor & 3U Eurocard & 160\,mm deep subrack. IEEE compliant. \\
        Expansion slots & 16, 7, or 4 & 19" and 10" rack compatible. \\
        Input power & 12-72\,V & Power directly from batteries. \\
        Communication & Ethernet & Packets are serialized with \textit{Protobuf} \cite{protobuf} over UDP-IP. \\
        \hline
        
        \rowcolor{gray!20} \multicolumn{3}{|c|}{\textbf{RTD Card}} \\
        \hline
        Channel count & 8 & 4-wire differential. \\
        Power consumption &  250\,mW & Requires split $\pm5$\,V supply. \\
        RTD resistance range & 0-5\,M$\Omega$  & \\ 
        Bias range & 500\,pA - 10\,$\mu$A & Pseudo-constant current bias. See section \ref{sec:rtd}. \\
        Bias range count & 16 & Distributed logarithmically. \\
        Lock-in frequency & 20 Hz & \\
        Bandwidth & 10\,Hz @ 80\,sps & \\
        \hline
        
        \rowcolor{gray!20} \multicolumn{3}{|c|}{\textbf{Diode Card}} \\
        \hline
        Channel count & 16 & Bias and differential readout. See section \ref{sec:diode}. \\
        Power consumption & 200\,mW & 3.3\,V single supply. \\
        Bias current & 1\,$\mu$A - 10\,mA  & Set with resistor, nominal 10\,$\mu$A. \\ 
        Readout range & 0-2.5\,V & Extra attenuation can be configured. \\
        Readout voltage noise & 6\,$\mu$V$_\textup{PP}$ 0.1-10\,Hz & Referenced to input. \\ 
        Bias current noise  & 100\,\textit{p}A$_\textup{PP}$ 0.1-10\,Hz & 100\,k$\Omega$ resistor. Johnson noise is negligible. \\ 
        Bandwidth & 10\,Hz @ 20\,sps &\\
        \hline
        
        \rowcolor{gray!20} \multicolumn{3}{|c|}{\textbf{Power distribution and Heater Card}} \\
        \hline
        Channel count & 16 / 8 & See section \ref{sec:heater} and \ref{sec:distribution}. \\
        Max voltage (low/high) & 5\,V / 72\,V  & High power voltage equal to system power input. \\
        Max current & 25\,mA / 250\,mA / 3\,A  & See text for configurations. \\
        Output resolution & 12 bits & Bandwidth limited at 20Hz. \\
        Current feedback resolution & 10\,$\mu$A & \\
        Current feedback offset & $\pm$100 $\mu$A & Trim in software.\\
        \hline
    \end{tabularx}
    \caption{$\tau$HK specifications. See respective sections for more details on the daughter cards.}
    \label{tab:overview}
\end{table*}

\section{System Architecture}

\begin{figure*}
\centering
\begin{circuitikz}

\ctikzset{multipoles/dipchip/width=1.6}
\draw (0,0) 
  node[dipchip, 
    num pins=18, 
    hide numbers,
    no topmark, 
    external pins width=0,
    scale= 1.25,
    ] (MCU) {};

\node[font=\small] at ([yshift=12pt]MCU.north) {\shortstack{MCU Card\\STM32H723}};

\node[right, font=\tiny] at (MCU.bpin 5){Ethernet};
\draw[-{Latex[length=4pt]}] (MCU.bpin 5) -- ++(-1.5,0) node[midway, above, font=\tiny]{\shortstack{Protobuf\\To Agent}};

\def\mbX{0.25};

\tikzset{demux 16/.style={
  muxdemux,
  muxdemux def={Lh=3, NL=1, Rh=4, NR=16, NB=0, w=1.5, square pins},
  scale=0.9
}}

\node [left, font=\tiny] at (MCU.bpin 16) {\shortstack{SPI CS\\8 bit}};
\node[demux 16, anchor=lpin 1, name=SPIMUX ] at ($(MCU.bpin 16)+(\mbX,0)$){};
\node [font=\tiny, rotate=90] at (SPIMUX) {\shortstack{Binary\\Decoder}};
\draw (MCU.bpin 16) -- (SPIMUX.lpin 1); 

\coordinate (bus start) at ($(MCU.bpin 18 -| SPIMUX.rpin 1)$);

\node [left, font=\tiny] at (MCU.bpin 18) {\shortstack{SPI\\8\,Mbit/s}};
\draw (MCU.bpin 18) -- (MCU.bpin 18 -| bus start) ; 

\node [left, font=\tiny] at (MCU.bpin 13) {\shortstack{I$^2$C\\400\,kbit/s}};
\node[demux 16, anchor=lpin 1, name=I2CMUX] at ($(MCU.bpin 13)+(\mbX,0)$) {};
\node [font=\tiny, rotate=90] at (I2CMUX) {\shortstack{2xTCA9548}};
\draw (MCU.bpin 13) -- (I2CMUX.lpin 1); 

\node [left, font=\tiny] at (MCU.bpin 11) {Clocks};
\draw (MCU.bpin 11)  -- (MCU.bpin 11 -| bus start) node[midway, above, font=\tiny]{};

\node [left, font=\tiny] at (MCU.bpin 10) {Power};
\draw (MCU.bpin 10) -- (MCU.bpin 10 -| bus start) node[midway, above, font=\tiny]{3.3\,V};

\draw[very thick, -{Latex[length=6pt]}] 
  (bus start) -- ++(0,-9) node[below right, font=\scriptsize, rotate=90] {Backplane bus};


\foreach \ycoor in {1,-3}
{
    
    \ctikzset{multipoles/dipchip/width=2}
    \draw (6.5,\ycoor) 
      node[dipchip, 
        num pins=12, 
        hide numbers,
        no topmark, 
        external pins width=0,
        scale= 1,
        ] (card) {};
    \node[font=\small] at ([yshift=6pt]card.north) {Daughter Card};
    
    \node [right, font=\tiny] at (card.bpin 1) {SPI \& 2xCS};
    \draw (card.bpin 1) -- (card.bpin 1 -| bus start); 
    \node [right, font=\tiny] at (card.bpin 2) {Clocks};
    \draw (card.bpin 2) -- (card.bpin 2 -| bus start);
    \node [right, font=\tiny] at (card.bpin 3) {Power};
    \draw (card.bpin 3) -- (card.bpin 3 -| bus start);
    \node [right, font=\tiny] at (card.bpin 4) {I$^2$C};
    \draw (card.bpin 4) to[short, -*] ++(-1.75,0) coordinate(jct) -- (card.bpin 4 -| bus start) ;
    
    \tikzset{demux 2/.style={
      muxdemux,
      muxdemux def={Lh=1.5, NL=1, Rh=2.5, NR=2, NB=0, w=1.25, square pins},
      scale=1,
      draw only right pins = {},
    }}
    \node[demux 2, anchor=center, name=expander] at ($(card.bpin 5)!0.5!(card.bpin 6)+(-0.75,0)$) {};
    \node [font=\tiny, rotate=90] at (expander) {\shortstack{I/O\\Expander}};
    \draw (expander.lpin 1) -- (expander.lpin 1 -| jct) -- (jct);
    
    \node [right, font=\tiny] at (card.bpin 5) {6 ID Bits};
    \draw (card.bpin 5) -- (card.bpin 5 -| expander.rpin 1);
    \node [right, font=\tiny] at (card.bpin 6) {2 Status LEDs};
    \draw (card.bpin 6) -- (card.bpin 6 -| expander.rpin 1);

    \def\dbH{4}
    \tikzset{db37 body/.style={
      muxdemux,
      muxdemux def={Lh=4.0, NL=0, Rh=4.5, NR=0, NB=0, w=0.3, square pins},
      scale=1,
      fill=black!50,
    }}

    \node [db37 body, anchor=center, name=connector] at (card.bpin 10){};
    \fill[black!50] ($(card.bpin 7)+(-0.2,-0.1)$) rectangle ($(card.bpin 7)+(0.2,0.1)$);
    \draw[fill=red] ($(card.bpin 7)+(-0.1,0)$) node[ocirc]{};
    \draw[fill=green] ($(card.bpin 7)+(+0.1,0)$) node[ocirc]{};

}

\foreach \i in {0,0.25,0.5}
{
    \draw[fill=black] ($(6.5,-5.25)+(0,-\i)$) node[ocirc]{};
}

\ctikzset{multipoles/dipchip/width=2.6666666666666}
\draw (0,-5) 
  node[dipchip, 
    num pins=10, 
    hide numbers,
    no topmark, 
    external pins width=0,
    scale= 0.75,
    ] (PSU) {};
    
\node[font=\small] at ([yshift=6pt]PSU.north) {Power Supply};

\node [left, font=\tiny] at (PSU.bpin 6) {3.3\,VA};
\draw (PSU.bpin 6) -- (PSU.bpin 6 -| bus start);
\node [left, font=\tiny] at (PSU.bpin 7) {3.3\,VD};
\draw (PSU.bpin 7) -- (PSU.bpin 7 -| bus start);
\node [left, font=\tiny] at (PSU.bpin 8) {5\,V};
\draw (PSU.bpin 8) -- (PSU.bpin 8 -| bus start);
\node [left, font=\tiny] at (PSU.bpin 9) {$\pm$5\,VA};
\draw (PSU.bpin 9) -- (PSU.bpin 9 -| bus start);
\draw (PSU.bpin 10) -- (PSU.bpin 10 -| bus start);

\node [left, font=\tiny] at (PSU.bpin 3) {Input 12-72\,V};
\draw [densely dashed] (PSU.bpin 3) to[short, -] ++(0.5,0) coordinate(jct);

\draw [densely dashed] (PSU.bpin 6 -| jct) to[short, -*] 
    (PSU.bpin 7 -| jct) to[short, -*] 
    (PSU.bpin 8 -| jct) to[short, -*]
    (PSU.bpin 9 -| jct) to[short, -]
    (PSU.bpin 10 -| jct);
    
\coordinate(open) at ($(PSU.bpin 8) + (-1.25,0)$);
\draw [densely dashed] (PSU.bpin 6 -| open) -- (PSU.bpin 6 -| jct); 
\draw [densely dashed] (PSU.bpin 7 -| open) -- (PSU.bpin 7 -| jct); 
\draw [densely dashed] (PSU.bpin 8 -| open) -- (PSU.bpin 8 -| jct); 
\draw [densely dashed] (PSU.bpin 9 -| open) -- (PSU.bpin 9 -| jct); 
\draw [densely dashed] (PSU.bpin 10) -- (PSU.bpin 10 -| jct);

\end{circuitikz}
\caption{Block diagram of client architecture for $\tau$HK. The agent software running on a computer connects and controls one or more clients over Ethernet. In each client, the microcontroller (MCU) acts as a hardware abstraction layer for up to 16 daughter cards that provide application-specific functionality. The agent software provides an application programming interface (API) for user interaction with the system.}
\label{fig:block_diagram}
\end{figure*}
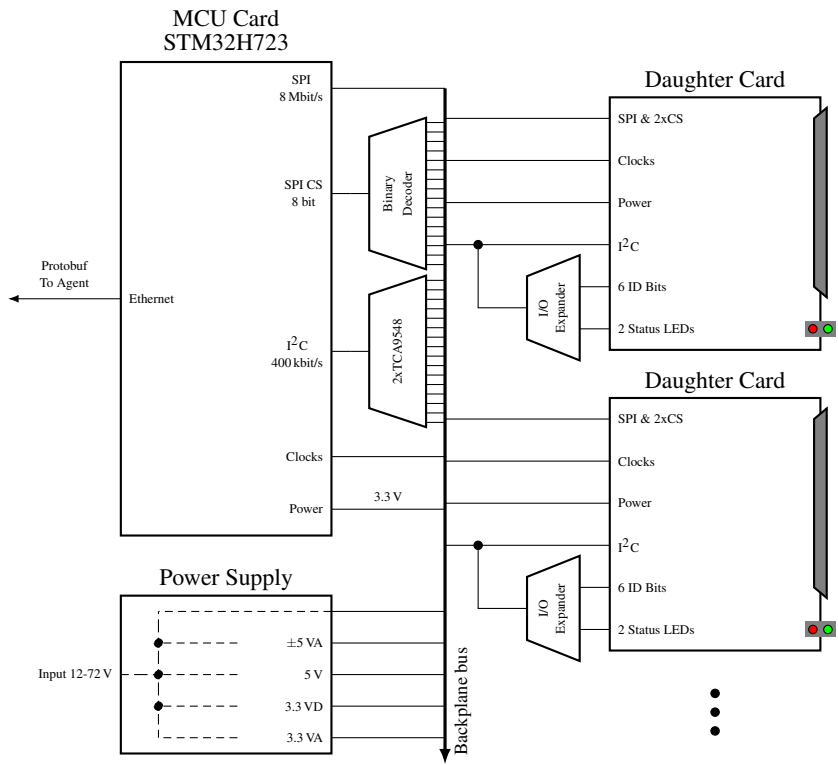

\tauHK uses an agent-client architecture, with a block diagram of the client presented in Fig. \ref{fig:block_diagram}. 
This design allows multiple clients to connect into a single computer, that aggregates incoming data and manages the state of all connected clients.
The agent software integrates with the \textit{Dirfile}\cite{getdata} and \textit{InfluxDB}\cite{influx} databases that pair with a variety of visualization tools such as \textit{KST}\cite{KST} and \textit{Grafana}\cite{grafana}. Commands can be sent to the system via a \textit{Python} library or using a graphical user interface served over the web.
Each client can accommodate up to 16 daughter cards, providing application-specific I/O capabilities, as seen in Fig. \ref{fig:picture}.

In addition, smaller backplanes were developed for mass- and space-constrained systems. 
Some of these systems may operate without a dedicated power supply as a single 3.3\,V supply has been integrated with the microcontroller.
The ability to distribute multiple small systems throughout an experiment reduces the complexity of custom wiring in favor of simple Ethernet cables.

A fully populated 16 card system can run 256 simultaneous lock-in amplifiers.
Traditionally, such a large synchronous load would require a field programmable gate array (FPGA) increasing complexity and cost.
However, \tauHK achieves this task using only the peripherals present on the STM32H723\cite{stm32} microcontroller.
Jitter-sensitive clocks are generated in hardware by the timer peripherals and low latency tasks are run inside interrupt service routines to never miss samples.
Using a generic microcontroller on \tauHK significantly reduces the cost and development time and simplifies firmware updates in the field. 
\tauHK was designed with cost and manufacturability in mind, the system pictured in Fig. \ref{fig:picture} costs under USD 2000\footnote{Purchased in 2024.} and can be assembled by most PCB assembly houses. 

\section{Daughter Cards}

The functionality of \tauHK is expanded by interchangeable, application-specific daughter cards that connect to the shared backplane.
Each card communicates with the microcontroller via a standardized set of interfaces, supplemented by a few global signals for clocking.
High bandwidth, timing sensitive data such as analog to digital converter (ADC) samples are transmitted over SPI running at 8Mbit/s with two dedicated chip select lines per card.
Slower control and configuration data is exchanged via \itwoc bus with each card having a unique name space provided by multiplexers on the backplane. 

The daughter cards are designed for ``plug-and-play'' operation as they requiring no runtime configuration.
Each card stores its unique calibration data onboard and is identified by the system with 6 hard-wired identification bits.
Most cards use a standardized front panel with a DB-37 connector for I/O and two status LEDs.

To streamline development and  reduce complexity of designing new cards, daughter cards share common design elements.
For example, the ADS131m\cite{ads131}
family of $\Sigma{-}\Delta$ analog to digital converter (ADC) is used on all existing cards.
Reusing hardware components reduces the code complexity and simplifies data synchronization across cards.

\subsection{RTD card}\label{sec:rtd}

\begin{figure*}
\centering
\begin{circuitikz}[american voltages]
\tikzstyle{every node}=[font=\large]

\def\xA{15}    
\def\xB{18.25} 
\def\xC{20.75} 
\def\xD{21.25} 
\def\xE{22.75} 
\def\xV{12} 

\def\yTop{13.75}  
\def\yPot{14}  
\def\yMid{12}  
\def\yBot{8.75}  
\def\yRt{11.25}   
\def\yRs{13.5} 
\def\yOp{10.0} 

\draw (\xB,\yBot) to[variable american resistor,l={$R_{RTD}$}] (\xB,\yRt);
\draw (\xA,\yRt) to[R,l={$R_L$}] (\xB,\yRt);
\node at (\xB,11.25) [circ] {};
\node at (\xB,8.75) [circ] {};

\draw (\xV, \yRt) to[R,l={$R_e(a)$}] (\xA, \yRt);


\draw (\xD,\yOp) node[op amp,scale=1, yscale=-1 ] (opamp2) {};
\draw (\xE-0.3,\yOp) to[short](\xE,\yOp);
\draw (\xE,\yOp) --++(0.2,0) node[above] {\small \textbf{ADC}} --++(0.2,0) [-latex];

\draw (\xB,8.75) to[short] (\xC-1,8.75);
\draw (\xC-1,9.5) to[short] (\xC-1,8.75);
\draw (\xC-1,9.5) to[short] (opamp2.-);
\draw (\xB,11.25) to[short] (\xC-1,11.25);
\draw (\xC-1,11.25) to[short] (\xC-1,10.5);
\draw (\xC-1,10.5) to[short] (opamp2.+);

\draw (\xV,\yRt) to[V,l={$V_e(a)$}] (\xV,\yBot);
\draw (\xV,\yBot) to[short] (\xB,\yBot);

\draw (\xB+1,\yRt) to [open, v^>=$V_b$] (\xB+1,9);

\draw[style={dashed, draw}] (\xV - 1, \yRt + 1)node[label={[align=center, anchor=west, yshift=4]above:{\small \textbf{Thévenin logDAC}}}]{} -- ++(4.5,0) -- ++(0,-4) --  ++(-4.5,0) -- cycle;

\end{circuitikz}
\caption{Simplified readout circuit -- the actual circuit features dedicated bias and sense lines and is fully differential. The RTD is placed in series with a $R_L=320$\,k$\Omega$ load resistor and the resulting voltage is amplified and measured by an ADC. The excitation is generated by a logarithmic attenuator (logDAC) as a function of the commanded attenuation $a$. The logDAC in $\tau$HK has 16 ranges and is characterized by $R_e(a)=2$\,k$\Omega$ and $V_e(a)=5V\cdot2^{-a}$ for $a\in[0,\dots,15]$.}
\label{fig:rtd_circuit}
\end{figure*}
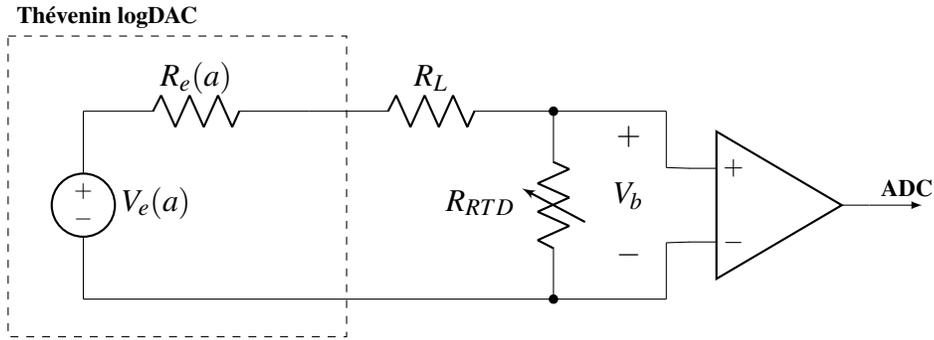

The readout scheme, shown in Fig. \ref{fig:rtd_circuit}, implements a 4-wire, pseudo-constant current bias with 16 software controlled ranges spanning 96\,dB logarithmically. 
The RTD is in series with a 320\,k$\Omega$ load resistor, and an instrumentation amplifier is used to measure the voltage across it to infer resistance.
To achieve the large dynamic range, each channel uses a discrete multiplying logarithmic digital to analog converter (logDAC) that attenuates the sine wave bias from a shared DAC.
The circuit is a fully differential lock-in amplifier to maximize common mode rejection and mitigate $1/f$ noise respectively.

The logDAC provides 16 discrete attenuation steps, each reducing the output by a constant factor of 2 while maintaining a constant output impedance. 
This logDAC is implemented as two symmetric resistor ladders to maintain the fully differential scheme each with 4 dual pole switches and 8 resistors.
Logarithmic resistor ladders are typically used for audio applications such as the Analog Devices \textit{DS1882} or Texas Instruments \textit{LM1971}, however these do not achieve the noise levels required in RTD readouts, thus necessitating a discreet solution.

The use of a common bias waveform reduces the computational load on the host processor as it does not need to compute unique amplitude sine waves for each channel.
Attenuating the excitation waveform also attenuates the noise produced by the shared DAC and reduces any common-mode offset errors.
The logDAC is comprised of analog switches (TMUX7234 \cite{TMUX7234}) that specify a small 0.1\,nA maximum leakage current however the scheme is fully symmetric so only a fraction of that current actually flows across the RTD.
Without any active common-mode reduction circuitry, the leakage is maximal for low resistance RTD as all the amplifier bias current flows across the RTD \cite{AD8421}.
For higher resistance RTDs, the input bias current is shunted back through the bias circuit with only a fraction contributing to common mode heating.

\begin{figure}
    \centering
    \includegraphics[width=\linewidth]{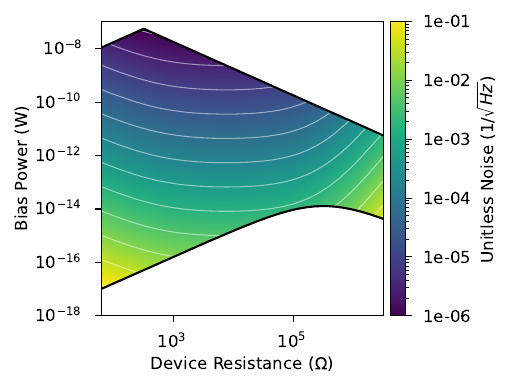}
    \caption{Fractional white noise level for noiseless resistor biased and readout by the RTD card from an empirical noise model. Measured performance matches or exceeds the presented data. The upper limit on bias is set by the maximal readout voltage and the lower limit is set by the maximal logDAC attenuation.}
    \label{fig:rtd_noise}
\end{figure}

The resistance noise power spectral density depends on the RTD's resistance and the applied bias power -- limited by self-heating.
Fig. \ref{fig:rtd_noise} presents the fractional white noise contribution of the readout and bias circuit for varying RTD resistances and bias powers.
Since cryogenic RTDs have dimensionless sensitivities on the order of 1, the fractional resistance noise can be interpreted in temperature units\cite{cernox}.

The primary noise contribution comes from the AD8421 \cite{AD8421} instrumentation amplifier connected to the sense lines coming from the RTD.
This noise contribution is independent on the bias circuit and therefore the fractional noise decreases as the bias is increased -- injecting more signal.
The maximum and minimum bias power is a function of device resistance due to the pseudo-constant current bias scheme.
Device resistances scale the current noise linearly and dominates the noise contribution for resistances above a few tens of k$\Omega$ resulting in worse performance for high resistance RTDs.
For high-impedance RTDs such as neutron transmutation doped germanium, the current noise may be mitigated with the addition of JFET followers\cite{blastbus}.

\begin{figure}
    \centering
    \includegraphics[width=\linewidth]{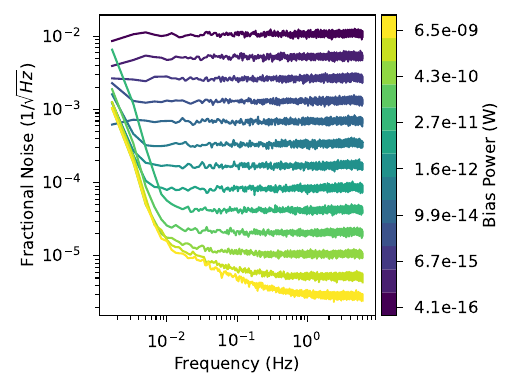}
    \caption{Noise spectral density measuring a 2\,k$\Omega$ room temperature resistor with 13 logDAC states. Larger bias powers introduce additional $1/f$ noise until the gained sensitivity outweighs the added noise and the $1/f$ noise plateaus at the limit set by the thermal environment and resistor under test.}
    \label{fig:rtd_low_f}
\end{figure}

The spectral shape of the noise is presented in Fig. \ref{fig:rtd_low_f}.
Due to their large temperature coefficient, the analog switches forming the logDAC inject low frequency noise into the system.
The logDAC is designed such that later attenuation stages reduce the noise produced by the previous ones and therefore the amount of noise injected depends on the commanded attenuation.
For very low bias powers this strategy buries the low frequency noise below the white noise threshold, however with larger powers there is less attenuation and therefore more noise makes it through.
For the 2\,k$\Omega$ resistor used to collect data for Fig. \ref{fig:rtd_low_f}, the low frequency noise peaks for the 8th logDAC step after which the increase in signal from the next logDAC step outweighs the increase in noise.

\subsection{Diode card}\label{sec:diode}

The 16 channel diode card provides a precision DC or chopped constant current source along with a high-impedance differential amplifier for general purpose bias and readout applications.
At 4\,K silicon diodes typically exhibit 400\,$\mu$K$_{P-P}$ of noise and room temperature PTC thermistors biased at 10\,$\mu$A typically show 20\,m$^\circ$C$_{P-P}$ noise in a thermally stable environment.
Detailed specifications and performance of the diode card are presented in Table \ref{tab:overview}.

Each channel is configurable in hardware and in software.
Two jumpers allow the user to disconnect the bias current, and the magnitude of the bias current is programmed with a single resistor.
This modularity enables the card to be used for a variety of use cases, such as configuring two channels together for 4-wire measurements, driving optical photo-interrupters, or reading out strain gauges.
All bias sources can be controlled in software, allowing them to be turned on or off as needed or switched into chopping mode.

Chopping readout was developed for applications with large intrinsic $1/f$ noise or when absolute DC precision is required, such as optical photo-interrupters and strain-gauges respectively. 
In chopping mode, the bias switches to a square wave that is phase locked to the RTD bias sine wave. 
Lock-in readout cannot mitigate all sources of $1/f$ noise particularly when used with highly nonlinear devices like diodes.
Any time varying current injected into the sensor, such as from soft shorts to other circuits, or AC magnetic/capacitive pickup, will not be improved by chopping mode.

\subsection{Cryo heater card}\label{sec:heater}

The cryogenic heater card provides 16 channels of hardware configurable and software controllable outputs designed for driving heaters in cryostats.
Each channel can be configured as low or high power with hardware defined defaults ensuring safe operation in the event of control failures.
High power channels can deliver up to 250\,mA at input line voltage with 12 bit, low frequency pulse width modulation (PWM) control phase locked with the RTD excitation waveform, while low power channels utilize 12 bit DACs to supply up to 25\,mA at 5\,V.
Both types of channels have over-current and short circuit protection and transition to high impedance when commanded \textit{off}.
Further details on the specifications are presented in Table \ref{tab:overview}.

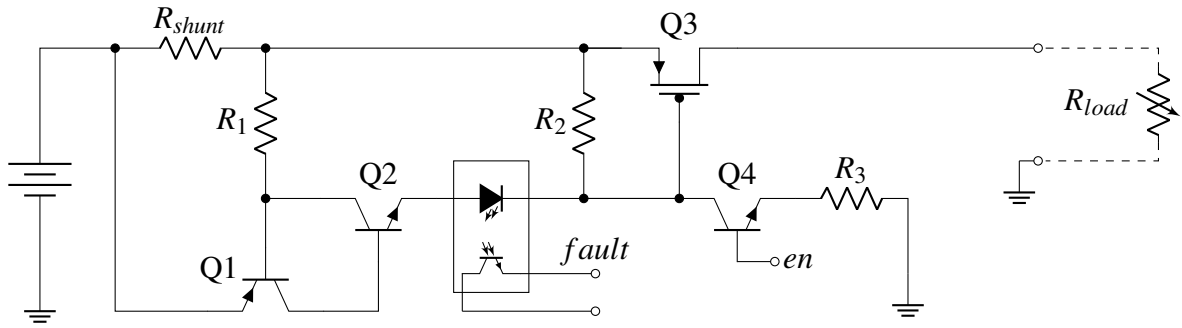
\begin{figure*}[t]
\centering
\begin{circuitikz}[american voltages]
\tikzstyle{every node}=[font=\large]

\ctikzset{tripoles/mos style/arrows}
\ctikzset{resistors/scale=0.7, capacitors/scale=0.7, inductors/scale=0.7}
\ctikzset{every node/.style={font=\small}}

\def\xRight{10}
\def\xLeft{1}
\def\xRone{3}
\def\xQtwo{4.5}
\def\xPmos{8.5}
\def\xOpto{6}

\def\yTop{10}
\def\yMid{8}
\def\yQone{6.5}
\def\yQtwo{8}

\draw node[pnp, rotate=-90, yscale=-1, label={[above left=10pt]Q1}](Q1) at (\xRone, \yQone) {};
\draw node[npn, rotate=90, label=Q2](Q2) at (\xQtwo, \yQtwo) {};
\draw (Q1.C) to[short, -] (Q1.C -| Q2.B) -- (Q2.B);

\draw (\xPmos,\yTop) node[pmos, rotate=90, label={[above]Q3}](Qp) {};


\draw (Qp.E) to[short, -*] ++(-0.5,0) coordinate(Rup)
to[R, -*, l_={$R_2$} ] (Q2.center -| Rup)
to[short, -*] (Q2.center -| Qp.B) -- (Qp.B);

\draw (Q2.center -| Qp.B) -- ++(0,0) node[npn, anchor=C, rotate=90, label={[above]Q4}](Qen){};
\draw (Qen.E) to[R, l={$R_3$}] ++(1.5,0) -- ++(0,-1) coordinate(GroundRight) node[ground]{};
\draw (Qen.B) to[short,-o] ++(0.5,0) node[right]{$en$};
\draw (\xRone, \yTop) -- (Qp.E);

\draw (Q2.E) -- (\xOpto-0.5, \yMid) to[leD*, diodes/scale=0.5, led arrows from cathode,mirror]  (\xOpto+0.5, \yMid) --(Q2.center -| Qp.B);
\draw node[npn,photo,rotate=-90, yscale=-0.5, xscale=0.5](Qphoto) at (\xOpto, \yMid-1){};
\draw (\xOpto-0.5, \yMid-1.25) rectangle (\xOpto+0.5, \yMid+0.5);

\draw (Qphoto.E) to [short, -o] ++ (1,0) coordinate(there) node[above]{$fault$};
\draw (Qphoto.C) -- ++ (0,-0.5) coordinate(here);
\draw (here) to[short, -o] (here-|there);


\draw (\xLeft, \yTop) to[american resistor, l={$R_{shunt}$}, -*] (\xRone, \yTop)
to[american resistor, l_=$R_1$] (\xRone, \yMid) -- (Q1.B);

\draw (Q2.C) to[short, -*] (\xRone, \yQtwo);

\draw (\xLeft,\yTop) coordinate(ShuntLeft) to[short, *-] (Q1.E-| ShuntLeft) -- (Q1.E);
\draw (ShuntLeft) -- ++(-1,0) to[battery] ++(0, -3.5) node[tlground]{};

\draw (Qp.C) to[short, -o] ++ (4,0) coordinate(hereCirc);
\draw ($(hereCirc)+(0,-1.5)$) to[short, o-] ++(-0.25,0) coordinate(hereGnd) node[ground]{};
\draw[dashed] ($(hereCirc)+(0.1,0)$) -- ++(1.5,0) -- ++(0,-0.25) coordinate(here);
\draw (here) to[vR, l_={$R_{load}$}] ++ (0,-1) coordinate(here);
\draw[dashed] (here) -- ++(0,-0.25) -- ++(-1.5,0);

\end{circuitikz}
\caption{High-side protection circuit used on high power heaters. When the current drawn by the load increases such that the voltage across $R_{shunt}$ exceeds the BE voltage of Q1, the transistor pair Q1 and Q2 latch on pulling up the gate of Q3 and turning off the channel. Q1 and Q2 along with a bypass network form a discrete silicon controlled rectifier (SCR) that once latched on can only be reset by toggling Q4 with the enable signal. Time to trip is set by the $RC$ time constant of $R_1\times\beta_{Q_1}$ and $C_{bypass}$ (omitted for clarity). The fault state, indicated by current flowing through the SCR, can be detected by the opto-isolator.}
\label{fig:heater_protection}
\end{figure*}

High power channels have a novel high side current limiting protection similar to fold-back protection but designed to latch off until reset.
As illustrated in Fig. \ref{fig:heater_protection}, this protection circuit is built around a discrete silicon controlled rectifier (SCR) that latches the output off when the voltage across a shunt exceeds a threshold.
The tripping point is determined by the shunt resistor and base-emitter voltage drop of the PNP transistor, a characteristic that varies greatly from part to part and over temperature.
To prevent premature tripping, the shunt resistor is chosen such that the tripping point is nominally 500\,mA for channels rated at 250\,mA.
Due to the fast response of the protection circuit, the heater card is not designed to   drive low impedance loads, even at reduced duty cycles that maintain an average current below the rated limit.
When choosing a suitable load impedance for heaters, the user must select a resistance that would not exceed the channel capacity when driven at full power\footnote{For a 48\,V input system the minimum impedance is 200\,$\Omega$}.

High power channels exclusively use PWM control, as linear regulation of 250\,mA across large voltage differentials would result in an unmanageable power dissipation at the desired channel density.
Switching noise is mitigated through the addition of Miller capacitance and a deliberately slow gate driver circuit limiting the output slew rate to 200\,V/ms.
With this mitigation in place, RTD and diode time-streams are immune to the transients generated by PWM switching for all wiring configurations tested.

Current feedback is implemented by monitoring the voltage across a low-side shunt resistor.
The heater card is not designed for high precision, so the current feedback exhibits some DC offset, and the shunt may show non-linearity at higher power levels.
The current time stream is digitally low pass filtered to a bandwidth of 10\,Hz.

\subsection{Power distribution card}\label{sec:distribution}

An 8 channel, fully isolated, power distribution card is used for general power switching applications such as powering sub-sections of the instrument or switching arbitrary large loads.
The power distribution card re-uses the high power section of the cryo heater card with some minor modifications aimed at increasing the current capacity from 0.25\,A to 3\,A.
Each channel exhibits less than 500\,m$\Omega$ ohmic series resistance and draws 0.3\,mA of standby current.

This card can operate in a remote mode where it takes on/off commands from open-collector parallel inputs.
The full truth table is presented in Table \ref{tab:power_switching_truth}.
Remote mode is useful for interoperability with stand-alone electronics such as the CSBF \textit{science stack}\cite{LDB_science_support_package} or physical buttons.

Unlike the other daughter cards, the power distribution card draws input power from a front panel connector as the maximum load current would overwhelm the backplane.
The output channels are fully isolated from the backplane and the card can still operate in remote mode when the rest of \tauHK is un-powered.
Powering \tauHK through this card is a convenient way to add a self hard-reset functionality to a system.
\captionsetup[table]{skip=2pt} 
\begin{table}
\renewcommand{\arraystretch}{1.3}
\centering
\begin{minipage}[t]{0.45\textwidth}
\centering
\vspace{0pt}
\caption*{\textbf{Remote Latch}}
\begin{tabularx}{\textwidth}{>{\centering\arraybackslash}X >{\centering\arraybackslash}X | >{\centering\arraybackslash}X}
\toprule
\rowcolor{gray!20} $\overline{\textbf{Set}}$ & $\overline{\textbf{Reset}}$ & \textbf{Latch} \\
\midrule
\fallingedge & 1 & 1 \\
1 & \fallingedge & 0 \\
1 & 1 & Last \\
\multicolumn{2}{c|}{\textit{Power-on Reset}} & 0 \\
\bottomrule
\end{tabularx}
\end{minipage}
\hfill
\begin{minipage}[t]{0.45\textwidth}
\centering
\vspace{10pt}
\caption*{\textbf{Channel Output}}
\begin{tabularx}{\textwidth}{>{\centering\arraybackslash}X >{\centering\arraybackslash}X | >{\centering\arraybackslash}X}
\toprule
\rowcolor{gray!20} \textbf{Latch} & \textbf{MCU} & \textbf{Output} \\
\midrule
0 & X & 0 \\
X & 0 & 0 \\
1 & 1 & 1 \\
1 & Off & 1 \\
\bottomrule
\end{tabularx}
\end{minipage}
    \caption{Truth table for remote operation of the power distribution card. The remote \textit{Set/Reset} inputs connected to external switches or open-collector outputs form an \textit{SR} latch. The latch state is combined with the microcontroller's output via a logical \textit{AND} operation. In case of microcontroller outages, the latch takes effect.}
    \label{tab:power_switching_truth}
\end{table}

\section{Conclusion}

\tauHK has been designed to meet the requirements of the Taurus balloon experiment but remains flexible enough to be used for many large scale experimental platforms.
The functionality of \tauHK can be expanded through the use of application-specific daughter cards communicating with the host over a standardized set of channels.
This paper presents a base set of expansion cards that cover the major use cases for most cryogenic experiment detailed in Table \ref{tab:overview}.
Data streams and user interactions use an application programming interface (API) running on an agent computer communicating with the hardware over Ethernet.

\tauHK has been employed over multiple cryogenic runs in several labs spanning over a year of testing. 
It has successfully undergone both representative vacuum chamber and beamline radiation testing, validating that it is a suitable option for scientific ballooning experiments.
Please reach out to \textit{simont@princeton.edu} if you would like to inquire about using \tauHK for your lab or upcoming experiment.

\begin{acknowledgments}
The development of \tauHK was supported by NASA award number 80NSSC21K1957.

\end{acknowledgments}

\bibliographystyle{plainnat}
\bibliographystyle{unsrtnat}
\bibliography{aipsamp}

\end{document}